\documentclass[12pt]{article}

\catcode`\@=11

\global\arraycolsep=2pt
\usepackage{amsbsy,amssymb,latexsym,amsfonts,amsmath}
\usepackage{graphicx,color}

\begin{document}
\begin{flushright}
\parbox{4.2cm}
{CALT-TH-2014-166}
\end{flushright}

\vspace*{0.7cm}

\begin{center}
{ \Large Perturbative search for dead-end CFTs}
\vspace*{1.5cm}\\
{Yu Nakayama}
\end{center}
\vspace*{1.0cm}
\begin{center}
{\it Walter Burke Institute for Theoretical Physics, California Institute of Technology,  \\ 
Pasadena, California 91125, USA}
\vspace{3.8cm}
\end{center}

\begin{abstract}
To explore the possibility of self-organized criticality, we look for CFTs without any relevant scalar deformations (a.k.a dead-end CFTs)
within power-counting renormalizable quantum field theories with a weakly coupled Lagrangian description. In three dimensions, the only candidates are pure (Abelian) gauge theories, which may be further deformed by Chern-Simons terms. In four dimensions, we show that there are infinitely many non-trivial candidates based on chiral gauge theories. Using the three-loop beta functions, we compute the gap of scaling dimensions above the marginal value, and it can be as small as $\mathcal{O}(10^{-5})$ and robust against the perturbative corrections. These classes of candidates are very weakly coupled and our perturbative conclusion seems difficult to refute. Thus, the hypothesis that non-trivial dead-end CFTs do not exist is likely to be false in four dimensions.

\end{abstract}

\thispagestyle{empty} 

\setcounter{page}{0}

\newpage

\section{Introduction}

The rules of the game are as follows:

\begin{itemize}
\item
We look for conformal field theories (CFTs) without any relevant scalar deformations. We name them dead-end CFTs.
\item
We do not ask what will happen after introducing relevant deformations (if any).
\item
We do not impose any continuous global symmetries or discrete global/gauge symmetries.\footnote{Otherwise, we would have scalars with a shift symmetry or fermions with a (discrete) chiral symmetry as trivial examples.} 
\item
We assume that dead-end CFTs are unitary, causal, and have finite energy-momentum tensors.\footnote{Otherwise, generalized free CFTs would be trivial examples.}
\item Deformations must be physical. In gauge theories, they must be BRST invariant.\footnote{Otherwise, ghost mass terms or gauge non-invariant mass terms  would give unphysical relevant deformations. \label{foot1}}
\item
(Optional) We assume the absence of gravitational anomalies.
\item
(Optional) We only discuss power-counting renormalizable weakly coupled Lagrangian field theories in three and four dimensions.\footnote{To the author's knowledge, there are no known non-perturbative examples other than AdS/CFT inspired ones. We have an example of a dead-end CFT in two dimensions. See section 4.}

\end{itemize}

Let's play!

\subsection{Physical background of the game}
This game is designed to examine the possibility of self-organized criticality \cite{Bak:1987xua} (see e.g. \cite{reviews} for a review) in quantum field theories. 
In statistical systems, it is typically the case that in order to obtain criticality, we have to tune at least one parameter of the system (e.g., temperature). It is interesting to see if we can construct a self-tuning model in which the criticality is automatically attained by  just making the size of the system larger without tuning anything else.
A naive guess is that unless we use some symmetry principles (e.g., a Nambu-Goldstone or anomaly cancellation mechanism) generic gapless systems are unstable, and self-organized criticality is difficult to achieve.

However, we know one example in nature: the theory of the photon. Maxwell theory is always at criticality and we cannot detune the theory to make it gapped (i.e., to give the photon a mass) unless we introduce extra light matter fields (like the Higgs mechanism). The fact that it is always at criticality led Einstein to the discovery of special relativity. The speed of light is absolute. It is hard to imagine if he could have come up with his various gedanken experiments if the photon were massive or if the propagation of light were not critical.
The criticality of the photon is not protected by any global symmetry. It is an intrinsic property of Maxwell theory that it does not allow any relevant deformations.\footnote{In the BRST quantization, one may regard photon as a Nambu-Goldstone boson for the residual gauge symmetry $\delta A_\mu = a_\mu$. Since there is no way to break this symmetry in a physical manner, this fact is not important for our discussions. See also footnote \ref{foot1}.} It is an example of a dead-end CFT.

Is this just a peculiar coincidence or a deep feature of particle physics in  the particular case of four space-time dimensions?

Putting philosophical questions aside, one technical reason for being interested in the (non-)existence of dead-end CFTs is whether we can regularize various infrared singularities in the ``S-matrix" of CFTs. Strictly speaking, the S-matrix does not exist in CFTs, but once they are deformed to a massive/gapped/topological phase, the concept makes sense. Indeed, a clever use of the (regularized) S-matrix and its analyticity properties has led to many important results in quantum field theories such as the proof of the $a$-theorem in four dimensions \cite{Komargodski:2011vj},  enhancement of scale invariance to conformal invariance  \cite{Polchinski:1987dy}\cite{Luty:2012ww}, convexity properties of large twist operators in general CFTs \cite{Komargodski:2012ek}, and so on. 

One crucial point in the argument of enhancement of scale invariance to conformal invariance is as follows: If the theory were scale invariant but not conformally invariant, the argument in \cite{Polchinski:1987dy}\cite{Luty:2012ww} suggests that the ``c"-function (or ``a"-function in four dimensions) would decrease forever along the RG flow. However, {\it if} the theory can be deformed to a massive/gapped/topological phase, the central charge is bounded $c\ge0$ (or $a\ge 0$), and hence there is a contradiction. This argument does not apply if the theory under consideration is a dead-end CFT or more precisely a dead-end scale-invariant field theory \cite{Nakayama:2013is}.

In this paper, we look for candidates for dead-end CFTs within power-counting renormalizable quantum field theories with a weakly coupled Lagrangian description. Of course, it is desirable to give a non-perturbative argument that does not rely on perturbation theory or a Lagrangian description. It is, however, sufficient to give a perturbative example if we would like to disprove the claim that dead-end CFTs do not exist. We will give some further thoughts on non-perturbative aspects of the game in section 4.

\section{No non-trivial candidates in $d=3$}
We begin with the matter content of renormalizable quantum field theories that have a weakly coupled Lagrangian description in three dimensions. It consists of a certain number of bosonic spin-zero scalar fields and fermionic spin-half spinor fields charged under gauge groups that have suitable kinetic terms.
 In the search for dead-end CFTs, we would like to first ask under which conditions these Lagrangian theories may or may not admit Lagrangian mass terms.\footnote{Beyond perturbation theory, the existence of Lagrangian mass terms does not always imply the existence of relevant deformations. For instance, the Konishi operator in  planar (i.e., in the large-$N$ limit) $\mathcal{N}=4$ super Yang-Mills theory becomes irrelevant in the large 't Hooft coupling limit. In the finite-$N$ case, the Konishi operator mixes with multi-trace operators so that one combination remains relevant. See \cite{Beem:2013qxa} for example.} 

We can always choose a  real basis for the scalar fields $\phi^I$, and then they transform as real (linear) representations under the gauge groups. The existence of the kinetic term means that there exists a positive definite bilinear form $g_{IJ}$ so that the kinetic term $g_{IJ} D_\mu \phi^I D^\mu \phi^J$ is gauge invariant and non-degenerate. 
One can use the same bilinear form to construct gauge invariant mass terms for the scalars proportional to $g_{IJ}\phi^I\phi^J$.  These mass terms are relevant deformations with the power-counting scaling dimension $\Delta = 1$.

In a similar way, we can always choose a  Majorana basis for the fermionic spinor fields $\psi^a$ in three dimensions such that they transform as real representations under the gauge groups. Again, the existence of the kinetic term implies that there exists a positive definite bilinear form $g_{ab}$ such that the kinetic term $g_{ab} \bar{\psi}^a \gamma^\mu D_\mu \psi^b$ is gauge invariant and non-degenerate. As in the scalar case, one can use the same bilinear form to construct gauge invariant Majorana mass terms (or real mass terms) for the fermions proportional to $g_{ab} \bar{\psi}^a \psi^b$. These are relevant deformations with the power-counting scaling dimension $\Delta = 2$.

Therefore, in three dimensions,  gauge theories with any matter admit relevant deformations, namely mass terms, irrespective of their representations under the gauge groups as long as the kinetic terms exist. Within a weakly coupled Lagrangian description, they cannot be candidates for dead-end CFTs because the mass terms are always (perturbatively) relevant.

The only remaining possibilities are pure gauge theories. 
There are two possible choices of the kinetic terms, i.e. Yang-Mills kinetic terms or  Chern-Simons terms. The latter is a little because the Lagrangian is not gauge invariant. In any case, the introduction of Chern-Simons terms makes the theory topological in the infrared so such theories are not candidates for dead-end CFTs. As for the Yang-Mills kinetic terms, in three dimensions, we believe that gauge theories with non-Abelian gauge groups confine with a mass gap (although we do not know the rigorous proof), so the infrared theories are massive, and they are not candidates for dead-end CFTs.
Therefore, the only remaining candidates are pure Abelian gauge theories with a Maxwell-type action. It is scale invariant but not manifestly conformally invariant (see e.g. \cite{Jackiw:2011vz}\cite{ElShowk:2011gz}), so it may be better to call such theories dead-end scale-invariant field theories. Since change of the gauge coupling constant is a marginal deformation, they do not possess any relevant scalar deformations  expressible as an integral of gauge-invariant local operators, but we may add Chern-Simons terms so that they become topological in the infrared.

To avoid misunderstanding, we would like to comment on the non-perturbative fixed point which was claimed to be an example of self-organized criticality in certain spin liquid systems in $d=1+2$ dimensions (see e.g. \cite{Hermele:2005}\cite{Hermele:2008afa} and references therein) with emergent Lorentz invariance. The effective field theories describing such spin liquids are given by (emergent) $U(1)$ gauge theories coupled with $N_f$ Dirac fermions (in the above Majorana basis we have used, a real vector representation of $O(2)$ gauge symmetry). In the large $N_f$ limit, the theories are supposed to be conformally invariant in the infrared. The crucial claim here is that all the relevant deformations such as fermion mass terms are forbidden by global symmetries such as $N_f$ flavor symmetries, parity, and time-reversal. While physically relevant, we do not consider them as examples of our dead-end CFTs because they violate the third rule  of the game.\footnote{Indeed, we can see that the introduction of a symmetry principle makes the better-known Banks-Zaks fixed point in four dimensions as an example of self-organized criticality. Given so many examples, there is less interest in pursuing such possibilities from the purely theoretical (zoological) viewpoint.}

\section{Non-trivial candidates in $d=4$}
We have seen that in three dimensions there are no non-trivial candidates for dead-end CFTs with a weakly coupled Lagrangian description. This section will show that the situation is drastically different in four dimensions because mass terms of fermions can be forbidden without using any global symmetries.

We start with the field content.
Renormalizable field theories in four dimensions with a weakly coupled Lagrangian description can have bosonic spin-zero scalar fields and fermionic spin-half spinor fields, charged under the gauge group, with finite kinetic terms. 
The argument for the scalars is the same as in three dimensions.
We can always choose a real basis for the scalar fields $\phi^I$, and they transform as real (linear) representations of the gauge groups. The existence of the kinetic term means that there exists a positive definite bilinear form $g_{IJ}$ so that the kinetic term $g_{IJ} D_\mu \phi^I D^\mu \phi^J$ is gauge invariant and non-degenerate. 
One can use the same bilinear form to construct gauge invariant mass terms for the scalars proportional to $g_{IJ}\phi^I\phi^J$.  These are relevant deformations with the power-counting scaling dimension $\Delta = 2$.

However, the situation is different for spinors. We can choose a Weyl basis of the fermions $\psi^a$ so that the representations of the gauge group are complex in general. The complex conjugate $\bar{\psi}_a$ (with the opposite chirality) transforms as the complex-conjugate representation of $\psi_a$. The existence of the Weyl kinetic term means that there exists a Hermitian bilinear form $\delta^{a}_{\ b}$ so that the kinetic term $\delta^{a}_{\ b} \bar{\psi}_a \sigma^\mu D_\mu \psi^b$ is gauge invariant and non-degenerate. The crucial difference here is that unlike in three dimensions, we cannot use the bilinear form $\delta^a_{\ b}$ to construct Lorentz-invariant mass terms because $\bar{\psi}_a$ and $\psi^a$ have different chiralities. The gauge theories with Weyl fermions in non-real representations  are called chiral gauge theories and since they do not (always) possess mass deformations, they are good candidates for dead-end CFTs.

Not every chiral gauge theory is consistent. They may have gauge anomalies. The anomaly cancellation conditions are well-known. For each gauge group, we require
\begin{align}
\sum_F \mathrm{Tr}(R_F^a\{R_F^b ,R_F^c\}) = 0 \ ,
\end{align}
where $R_F$ is the representation matrix and the sum is taken over all the Weyl fermions. 
 Note that the condition is linear in the matter representation, so we can add anomaly-free matter combinations and the result is still anomaly free.
We only consider anomaly free-gauge theories.

Extreme examples are pure gauge theories. They do not have any matter at all, and we cannot add any mass terms for the gauge bosons by hand. However, we believe that non-Abelian gauge theories in four dimensions confine and have a mass gap. Therefore, they are not candidates for dead-end CFTs. On the other hand, pure Abelian gauge theories are perfectly good examples of dead-end CFTs. The gauge coupling constant is a marginal deformation and they do not possess any relevant deformations at all. Indeed, we know that the standard model ends up with the free Maxwell theory in the far infrared, and it is a dead-end CFT!
Are there any other non-trivial examples? This is what we want to pursue in the rest of this section.

Given the above discussion, the non-trivial candidates we have in mind are anomaly-free chiral gauge theories without any scalar fields. Classically, these candidates are all conformally invariant and their gauge coupling constants are marginal. Renormalization makes the gauge coupling constants run, and the question is whether there are non-trivial zeros of the beta functions of the gauge coupling constants. The answer depends on the details of the gauge groups and the fermion representations.
If the fixed points are infrared stable, all the gauge coupling constants are irrelevant, and there are no relevant deformations at the fixed point. These fixed points are dead-end CFTs.\footnote{Within perturbation theory, there is no candidate for the Virial current, so the fixed point is conformally invariant rather than merely scale invariant (see e.g. \cite{Nakayama:2013is} and reference therein for more details).}
This leads to the issue of conformal windows in chiral gauge theories.
 Rather than trying to determine the boundary of the conformal windows, our strategy is to find infinitely many examples of non-trivial zeros of the beta functions in which the perturbative computation of the beta functions (up to three loop order in this paper) is reliable. 

One comment on renormalizability is in order. One may ask if the chiral gauge theories  we will discuss are really renormalizable. At least within the power-counting renormalization, they are proved to be renormalizable, and certainly we are able to compute the physical observables in these CFTs at three loop order. Our examples will turn out to be no more exotic than the standard model as chiral gauge theories, and if we doubt their renormalizability (or realizability in nature), we should ask the same question about the standard model. See e.g. \cite{Luscher:2000hn} and references therein for further discussions on the non-perturbative renormalizability.

\subsection{Simple quiver-type chiral gauge theories}\label{quiver}
The easiest way to solve the anomaly cancellation condition is to study $SU(N_c)^K$ quiver-type gauge theories. The matter Weyl fermions  are in  the bifundamental representations of  adjacent gauge groups and represented by arrows. When the number of incoming arrows and outgoing arrows are the same at each node that represents  a simple gauge group, the theory is anomaly free. In order to forbid fermion mass terms, it is sufficient to make the directions of the arrows only one way between any pair of nodes.

For simplicity, we focus on the circular quiver gauge theories of $SU(N_c)^K$ with $N_f$ generations of bifundamental Weyl fermions\footnote{The conformal window of the model was also discussed in \cite{Poppitz:2009uq}.}:
\begin{align}
\cdots \overset{\times N_f}{\longrightarrow} SU(N_c)_1  \overset{\times N_f}{\longrightarrow} SU(N_c)_2  \overset{\times N_f}{\longrightarrow} \cdots  \overset{\times N_f}{\longrightarrow}  SU(N_c)_K  \overset{\times N_f}{\longrightarrow} SU(N_c)_1 \overset{\times N_f}{\longrightarrow}  \cdots
\end{align}
The beta functions of the system can be computed up to three loops by using the recent results reviewed in Appendix. 
The three-loop beta functions in the Modified Minimal Subtraction scheme are given by
\begin{align}
\beta_i &= \frac{g_i^3}{(4\pi)^2} \left[-\frac{11}{3} N_c + \frac{2}{3} N_f N_c  \right] \cr
&+\frac{g_i^3}{(4\pi)^2} \left[ \frac{g^2_i}{(4\pi)^2} \left\{-\frac{34}{3}N_c^2 + N_c N_f \left(\frac{10}{3}N_c + \frac{N_c^2 -1}{N_c} \right)  \right\} \right. \cr
&+ \left. \left(\frac{g_{i-1}^2}{(4\pi)^2} +\frac{g_{i+1}^2}{(4\pi)^2} \right) N_c N_f \frac{N_c^2-1}{2N_c}  \right]  \cr
& + \frac{g_i^5}{(4\pi)^4} \left[ \frac{g_i^2}{(4\pi)^2} \left\{-\frac{2857}{54}N_c^3 +  N_cN_f \left(\frac{1415}{54}N_c^2 + \frac{205}{18}N_c\frac{N_c^2-1}{2N_c} -\left(\frac{N_c^2-1}{2N_c}\right)^2 \right)  \right. \right. \cr
&- \left. \left.  N_c^2 N_f^2 \left(\frac{79}{54}N_c+\frac{11}{9}\frac{N_c^2-1}{2N_c}  \right) \right\} \right. \cr
&\left. + \left( \frac{g_{i-1}^2}{(4\pi)^2} + \frac{g_{i+1}^2}{(4\pi)^2} \right) N_f 2\left(2N_c - \frac{N_c^2-1}{2N_c} \right) \frac{N_c^2-1}{2N_c} \right] \cr
 &+ \frac{g_i^3}{(4\pi)^2} \left[\left(\frac{g_{i-1}^4}{(4\pi)^4}+ \frac{g_{i-1}^4}{(4\pi)^4} \right)\left\{ N_f \left(\frac{133}{18} N_c -  \frac{N_c^2-1}{2N_c}  \right)  \frac{N_c^2-1}{2N_c} \frac{1}{2} \right. \right. \cr
& \left. \left. -2 N_f^2 \frac{11}{9}   \frac{N_c^2-1}{2N_c} \frac{1}{2} \frac{1}{2} N_c \right\} \right]  \ .
\end{align}
for each gauge coupling constant $g_i$ ($i= 1,2\cdots,K$). Asymptotic freedom requires $N_f < 5.5$.\footnote{We may relax the condition of asymptotic freedom, but in practice we cannot find any additional weakly coupled  fixed points by relaxing the condition.} 
In order to obtain a weakly coupled fixed point, it is desirable that $N_f$ is close to the upper boundary of the asymptotic-freedom limit, so our main focus will be  $N_f= 5$.

We look for zeros of the beta functions. When $N_f^*<N_f<5.5$ with a certain critical number $N_f^*$, the zeros of the beta functions correspond to infrared stable fixed points, and we obtain good candidates for dead-end CFTs.
Once we find the zero of the beta functions, we can compute the anomalous dimensions of the field-strength operators $\mathrm{Tr}_i (F_{\mu\nu} F^{\mu\nu})$ from the Hessian matrix $\partial_i \beta^j|_{g_i = g_i^*}$. 
Up to three-loop order, the beta functions of the gauge coupling constants do not depend on the number of nodes  $K$ in the quiver. This is because we need at least $K$ fermion loops to obtain non-trivial $K$ dependence in the beta functions. 
On the other hand, the anomalous dimensions of the field strengths do depend on $K$ because we have to diagonalize the $K\times K$ Hessian matrix.

In principle, we also need to study the CP odd operators  $\mathrm{Tr}_i (\epsilon_{\mu\nu\rho\sigma} F^{\mu\nu} F^{\rho\sigma})$ with their coupling constants $\theta_i$ as theta terms beyond perturbation theory. Actually, 
$K-1$ of the $K$ theta terms are redundant operators in this theory because they can be removed by (anomalous) phase rotations of the Weyl fermions. The overall theta term, however, can be non-trivial. In perturbation theory nothing depends on this overall theta parameter. We do not know if the theta term is non-perturbatively renormalized or whether it will affect the beta functions. In any case, if we have an infrared fixed point, the anomalous dimension must be positive and our discussions are still valid.
In the other examples that we discuss in later subsections, all the theta terms are redundant operators in the action.

In our perturbative search, we may set $g_1 = g_2 = \cdots = g_K$. We find that  the other fixed points make some of the gauge coupling constants vanish, so we end up with effectively decomposed non-circular quivers.
We have listed the two-loop and three-loop anomalous dimensions of the permutation symmetric field strength $\sum_i \mathrm{Tr}_i F_{\mu\nu} F^{\mu\nu}$ for small values of $N_c$ in table \ref{table1}. They do not depend on the number of nodes $K$. The anomalous dimensions of permutation non-symmetric field strengths do depend on $K$. For example, if $N_c=3$, $N_f=5$, $K=3$, we have the eigenvalues
\begin{align}
(0.0155313, 0.00919003, 0.00919003) \ 
\end{align}
at three-loop order.
For $N_c=3$, $N_f=5$, $K=4$, we have the eigenvalues
\begin{align}
(0.0155313, 0.0113038, 0.0113038, 0.00707626) \ .
\end{align}
For $N_c=3$, $N_f=5$, $K=5$, we have the eigenvalues
\begin{align}
(0.0155313, 0.0126102, 0.0126102, 0.00788365, 0.00788365) \ ,
\end{align}
and so on. In every cases, all of the eigenvalues are positive, meaning that the fixed points are infrared stable.

\begin{table}[htbp]
\begin{center}
\begin{tabular}{|c|c|c|c|}
\hline
 $N_c$ & $N_f$ & 2-loop  &  3-loop   \\ \hline \hline
 3 & 5 &0.01563 &0.0155313 \\ \hline
 5 & 5 & 0.01488 & 0.0148063
 \\ \hline
3& 4 & 0.220
 & 0.203393  \\ \hline
5 & 4 & 0.207 & 0.193566
 \\ \hline
3&  3 & 1.39  & 0.978207
 \\ \hline
5 & 3 & 1.26
 & 0.930279
 \\ \hline
\end{tabular}
\end{center}
\caption{The anomalous dimension of the permutation-symmetric field strength of the $SU(N_c)$ chiral quiver gauge theories with $N_f$ generations of bifundamental Weyl fermions.  Each entry has an additional integer label $K\ge3$.}  \label{table1}
\end{table}

Although the beta functions are renormalization-group scheme dependent, the anomalous dimensions at the fixed point are physical quantities, and they do not depend on the choice of the renormalization scheme. Also note that the smallness of the coupling constant $g_i$ at the fixed point itself is not important because the physical expansion parameters can be different (e.g. the 't Hooft coupling $g_i^2 N_c$ may be more relevant). The ratio between the two-loop and three-loop predictions is  a good barometer whether or not the perturbation theory is reliable (assuming there is no accidental cancellation).

 It turns out that in all the examples we have studied, the three-loop contributions to the anomalous dimensions makes their values smaller than the two-loop predictions. We find that the loop expansion is not terribly bad for the anomalous dimensions of permutation-symmetric field strengths for $N_f=5$, where the ratio between the three-loop contribution and the two-loop contribution is at the percent order.
For comparison, we show the two-loop and three-loop anomalous dimensions of the Banks-Zaks fixed point \cite{Caswell:1974gg}\cite{Banks:1981nn} of $SU(N_c)$ gauge theory with $n_f$ Dirac fermions in the fundamental representation in table \ref{table2}.

We find that the structure of the beta functions of our chiral quiver with $N_f$ generations of Weyl fermions in the bifundamental representations is more or less similar to that of the Banks-Zaks theory with $n_f  = N_f N_c$ Dirac fermions in  the fundamental representation. The only difference at the two-loop level is that we have  twice as many contributions to the wave-function renormalization factors of fermions, which makes the fixed-point coupling smaller in our chiral quiver gauge theories than at the Banks-Zaks  fixed point. 
It is generically believed that the Banks-Zaks theory with $n_f = 5N_c$ Dirac fermions in fundamental representations are safely in the conformal window, so it is plausible (although not proved) that our chiral quiver gauge theories with $N_f = 5$ and any $N_c$ are in the conformal window as well. If this is the case, we have infinitely many classes of dead-end CFTs labelled by $N_c$ and $K$. The $K$ dependence of the anomalous dimensions is very small, but we recall that the number of these slightly irrelevant deformations (gauge kinetic terms) is given by $K$ and the operator contents are different.

\begin{table}[htbp]
\begin{center}
\begin{tabular}{|c|c|c|c|}
\hline
 $N_c$ & $n_f$ & 2-loop  &  3-loop   \\ \hline \hline
 3 & 16 &0.0022075&0.00220301 \\ \hline
 3 & 15 &0.02272 &0.022307 \\ \hline
3 & 12 & 0.36
 & 0.296 \\ \hline
5 & 27 & 0.0007501 & 0.000749578
 \\ \hline
5 & 25 & 0.02192 & 0.021558
 \\ \hline
5 & 20 & 0.34 & 0.285
 \\ \hline
\end{tabular}
\end{center}
\caption{The anomalous dimension of the field strength at the Banks-Zaks fixed point for $n_f$ Dirac fermions in the fundamental representation.}  \label{table2}
\end{table}

The lower values of $N_f$ may admit more strongly coupled dead-end CFTs.  For example, $N_f=4$ $SU(3)$ chiral quiver gauge theories can be compared with $SU(3)$ Banks-Zaks theory with $n_f = 12$ Dirac fermions in the fundamental representation. Recent lattice simulations seem to  more or less agree that the latter is in the conformal window (see e.g. \cite{Appelquist:2011dp}\cite{DeGrand:2011cu}\cite{Aoki:2012eq}\cite{Cheng:2013eu}\cite{Ishikawa:2013tua} and reference therein), and they suggest that our chiral quiver gauge theories with $N_f=4$ are also in the conformal window. Even $N_f=3$ $SU(3)$  chiral quiver gauge theories can be compared with the $SU(3)$ Banks-Zaks theory with $n_f=9$ Dirac fermions in the fundamental representation. The latter may possess a fixed point (with some controversies in the lattice simulations), which suggests that the former may also have a fixed point. 
The analysis based on the existence of topological excitations in \cite{Poppitz:2009uq}, however, predicts (but does not prove) that chiral quiver gauge theories have a smaller conformal window than  vector-like Banks-Zaks theories, and $N_f=4$ might have  been excluded already.
It would be interesting to determine the conformal window, but this is not the main purpose of our paper. We only attempt to offer evidence for the existence of dead-end CFTs so we are more interested in the weakly coupled fixed points. To our knowledge, there are no arguments that $N_f=5$ chiral quiver gauge theories, which are in the perturbative regime, lie outside of the conformal window.

\subsection{Anomaly-free chiral matter}\label{GGBY}
A more non-trivial way to obtain anomaly-free chiral gauge theories is to use the cancellation among various matter representations of Weyl fermions in gauge/gravitational anomalies. Particularly well-known matter combinations that cancel the anomaly for an $SU(N_c)$ gauge group is a generalized Georgi-Glashow model with one anti-symmetric representation and $N_c-4$ anti-fundamental representations \cite{Georgi:1974sy}, and a generalized Bars-Yankielowicz model with one symmetric representation and $N_c+4$ anti-fundamental representations \cite{Bars:1981se}.

We may generically consider an $SU(N_c)$ gauge theory with $N_a$ generalized
 Georgi-Glashow multiplets and $N_s$ generalized Bars-Yankielowicz multiplets.
In this subsection, we focus on a single gauge group and we discuss the quiver generalization in the next subsection. We remark here that the $N_s=3$ and $N_c=5$ model is the $SU(5)$ grand unified extension of the standard model (without the Higgs field). In fact, all of these chiral gauge theories were introduced as  models for particle physics.

From the formula in Appendix, the three-loop beta functions are computed as 
\begin{align}
\beta &= \frac{g^3}{(4\pi)^2} \left[-\frac{11}{3} N_c + \frac{2}{3} N_a(N_c-3) + \frac{2}{3}N_s(N_c+3)  \right] \cr
&+\frac{g^5}{(4\pi)^4} \left[ -\frac{34}{3}N_c^2 + \right. \cr
& \left. + N_a \left\{\left(\frac{10}{3} N_c + 2\frac{(N_c+1)(N_c-2)}{N_c} \right)\frac{N_c-2}{2} +(N_c-4) \left(\frac{10}{3}N_c + 2 \frac{N_c^2-1}{2N_c}\right)\frac{1}{2} \right\} \right.  \cr
&\left.  \left. + N_s \left\{\left(\frac{10}{3} N_c + 2\frac{(N_c-1)(N_c+2)}{N_c} \right)\frac{N_c+2}{2} +(N_c+4) \left(\frac{10}{3}N_c + 2 \frac{N_c^2-1}{2N_c}\right)\frac{1}{2} \right\} \right] \right. \cr
& + \frac{g^7}{(4\pi)^6} \left[ -\frac{2857}{54}N_c^3   \right. \cr
&+   N_a \left\{\frac{1415}{54}N_c^2 + \frac{205}{18}N_c\frac{(N_c+1)(N_c-2)}{N_c} -\left(\frac{(N_c+1)(N_c-2)}{N_c}\right)^2 \right\} \frac{N_c-2}{2} \cr
&+ N_s \left\{\frac{1415}{54}N_c^2 + \frac{205}{18}N_c\frac{(N_c-1)(N_c+2)}{N_c} -\left(\frac{(N_c-1)(N_c+2)}{N_c}\right)^2 \right\} \frac{N_c+2}{2} \cr
& + \left(N_a(N_c-4) + N_s(N_c+4)\right)\left(\frac{1415}{54} N_c^2 + \frac{205}{18} N_c \frac{N_c^2-1}{2N_c} - \left(\frac{N_c^2-1}{2N_c}\right)^2 \right) \frac{1}{2} \cr
& -N_a^2 \left( \frac{79}{54} N_c + \frac{11}{9} \frac{(N_c + 1) (N_c - 2)}{N_c}  \right) \left(\frac{N_c - 2}{2} \right)^2    \cr
& - N_a N_s \left(\frac{79}{54} N_c + \frac{11}{9} \frac{(N_c+1)(N_c-2)}{N_c}\right)\left(\frac{(N_c-2)}{2 } \frac{(N_c+2)}{2}\right) \cr
&-N_a (N_a(N_c-4)+N_s(N_c+4))\left(\frac{79}{54} N_c + \frac{11}{9} \frac{(N_c+1)(N_c-2)}{N_c}\right)\left(\frac{N_c-2}{2 } \frac{1}{2}\right) \cr
& - N_s N_a \left(\frac{79}{54} N_c + \frac{11}{9} \frac{(N_c-1)(N_c+2)}{N_c}\right)\left(\frac{(N_c-2)}{2 } \frac{(N_c+2)}{2}\right) \cr
& - N_s^{2 }\left(\frac{79}{54} N_c + \frac{11}{9} \frac{(N_c-1)(N_c+2)}{N_c}\right)\left(\frac{N_c+2}{2 }\right)^2 \cr
& - N_s (N_a(N_c-4)+N_s(N_c+4)) \left(\frac{79}{54} N_c + \frac{11}{9} \frac{(N_c-1)(N_c+2)}{N_c}\right)\left(\frac{N_c+2}{2 } \frac{1}{2}\right) \cr
& - (N_a(N_c-4)+N_s(N_c+4))N_a \left(\frac{79}{54}N_c + \frac{11}{9} \frac{N_c^2-1}{2N_c}\right) \left(\frac{N_c-2}{2 } \frac{1}{2}\right) \cr
& - (N_a(N_c-4)+N_s(N_c+4))N_s\left(\frac{79}{54} N_c + \frac{11}{9} \frac{N_c^2-1}{2N_c}\right) \left(\frac{N_c+2}{2 } \frac{1}{2}\right) \cr
& \left.- (N_a(N_c-4)+N_s(N_c+4))^2\left(\frac{79}{54} N_c + \frac{11}{9} \frac{N_c^2-1}{2N_c}\right) \left(\frac{1}{2 } \frac{1}{2}\right) \right] \ .
\end{align}
As mentioned, the theta term in these models is redundant, so we only have to consider the non-trivial zero of the gauge coupling constant.

We can now play the game of finding  very weakly coupled fixed points by changing $N_c$, $N_a$ and $N_s$.\footnote{We can find the two-loop discussions for $N_s N_a=0$ case in \cite{master}. When $N_s N_a=0$, \cite{Poppitz:2009uq} also gives an estimate of the conformal window based on the topological excitations. The latter claims that their's is the first estimate of the conformal window of these models. Apparently, the existence of non-trivial conformal fixed points in chiral gauge theories have not been studied much in the literature.} 
For example, when $N_c=5$, we present the most weakly coupled fixed point for a given fixed value of $N_s$ together with the anomalous dimension of the field strength in table \ref{table3}. We see that these theories are more weakly coupled than the $SU(3)$ Banks-Zaks fixed point theory with $n_f=15$ Dirac fermions in  the fundamental representation (see table \ref{table2}).
We also see that the difference between the two-loop prediction and the three-loop prediction is of the order of a percent, so the perturbation theory seems fairly reliable.

\begin{table}[htbp]
\begin{center}
\begin{tabular}{|c|c|c|c|c|}
\hline
 $N_c$ &$N_s$ & $N_a$ & 2-loop  &  3-loop   \\ \hline \hline
 5 & 0 &13  & 0.00622 & 0.006194  \\ \hline
 5 & 1 & 9 & 0.00607 & 0.006046  \\ \hline
5 & 2 & 5 & 0.00592
 & 0.005904  \\ \hline
5 & 3 & 1 & 0.00579 & 0.0057688 \\ \hline 
\end{tabular}
\end{center}
\caption{The anomalous dimension of the field strength in weakly coupled chiral $SU(5)$ gauge theories with $N_s$ generations of Bars-Yankielowicz multiplets and $N_a$ generations of Georgi-Glashow multiplets.}  \label{table3}
\end{table}

We may further investigate much more weakly coupled fixed points. In table \ref{table4}, we show the very weakly coupled fixed points for which the anomalous dimension of the field strength is smaller than that of $SU(3)$ Banks-Zaks fixed point with $n_f=16$ Dirac fermions in the fundamental representation. For another comparison, we also note that the anomalous dimension of the field strength of photons in QED at the scale of the electron mass is 
$\partial_\alpha \beta_\alpha|_{\alpha = \frac{1}{137}} = \frac{4\alpha}{3\pi}|_{\alpha = \frac{1}{137}} + O(\alpha^2)  \sim 0.003$ and also comparable.

We can see that some of these examples, such as $SU(35)$ with $N_s=0$, $N_f=6$, are extremely weakly coupled. Their anomalous dimensions are $10^{-2}$ times smaller than that of QED and so are their loop corrections. It is hard to imagine that the conclusion that these models have non-trivial conformal fixed points will be refuted by any other methods. Since the loop suppression is very large, we do not have to worry about the scheme dependence of the beta function at higher-loop order, either.

\begin{table}[htbp]
\begin{center}
\begin{tabular}{|c|c|c|c|c|}
\hline
 $N_c$ &$N_s$ & $N_a$ & 2-loop  &  3-loop   \\ \hline \hline
 9 & 0 &8   & 0.001762  & 0.00175977 \\ \hline
 9 & 1 & 6 & 0.0017432& 0.0017414 \\ \hline
13 & 0 & 7 & 0.0008206 & 0.00082021  \\ \hline
11 & 2 & 4 & 0.000123063  & 0.000123055 \\ \hline 
23 & 1 & 5 & 2.79044 $\times 10^{-5}$  & 2.79039  $\times 10^{-5}$  \\ \hline 
35 & 0 & 6 & 1.20187 $\times 10^{-5}$  & 1.20186 $\times 10^{-5}$ \\ \hline 
7 & 3 & 2 & 0.00030422   & 0.000304164 \\ \hline 
19 & 4 & 1 & 4.05191$\times 10^{-5}$  & 4.05182  $\times 10^{-5}$  \\ \hline 
31 & 5 & 0 & 1.51642 $\times 10^{-5}$  & 1.51641 $\times 10^{-5}$ \\ \hline 
9 & 4 & 0 & 0.0016901  & 0.00168861
 \\ \hline 
\end{tabular}
\end{center}
\caption{ Examples of extremely small anomalous dimensions of the field strength  in $SU(N_c)$ chiral gauge theories with $N_s$ generations of Bars-Yankielowicz multiplets and $N_a$ generations of Georgi-Glashow multiplets.} \label{table4}
\end{table}

\subsection{Quivers with external matter}\label{mixed}
One may wonder if the extremely weakly coupled examples presented in subsection  \ref{GGBY} are isolated exotic examples. Unlike the Banks-Zaks fixed point with Dirac fermions in the fundamental representation, there is no Veneziano limit \cite{Veneziano:1979ec} that produces infinitely many arbitrarily weakly coupled fixed points in a controllable manner.
Nevertheless, we would like to show that there exist infinitely many such examples of (numerically) very weakly coupled dead-end CFTs by combining the quiver constructions in section \ref{quiver} and the non-trivial chiral multiplets in section \ref{GGBY}. 

We study $SU(N_c)^K$ quiver gauge theories with $N_f$ generations of Weyl fermions (arrows between nodes) in the bifundamental  representations.
 Again, for simplicity, we consider the circular quiver.
In addition, at each nodes we add $N_a$ copies of generalized  Georgi-Glashow multiplets and $N_s$ copies of generalized Bars-Yankielowicz multiplets. 
\begin{align}
\begin{array}{ccccccccc}
& &N_a \mathrm{GG}&  & N_a \mathrm{GG} & &N_a \mathrm{GG} & &\\
& & \uparrow& \ & \uparrow & \ & \uparrow && \\
\cdots & \overset{\times N_f}{\longrightarrow} & SU(N_c)_K & \overset{\times N_f}{\longrightarrow} & SU(N_c)_1 &\overset{\times N_f}{\longrightarrow} &SU(N_c)_2 & \overset{\times N_f}{\longrightarrow} & \cdots \\
&& \downarrow& \ &\downarrow & \ & \downarrow && \\
&&N_s \mathrm{BY} & & N_s \mathrm{BY} & &N_s \mathrm{BY} && \\
\end{array}
\end{align}
The model is chiral and dose not admit any mass term.

The two-loop beta function at each node is given by
\begin{align}
\beta_i &= \frac{g_i^3}{(4\pi)^2} \left[-\frac{11}{3} N_c + \frac{2N_a}{3}(N_c-3) + \frac{2N_s}{3}(N_c+3) +  \frac{2}{3} N_f N_c  \right] \cr
&+\frac{g_i^3}{(4\pi)^2} \left[ \frac{g^2_i}{(4\pi)^2} \left\{-\frac{34}{3}N_c^2 + N_c N_f \left(\frac{10}{3}N_c + \frac{N_c^2 -1}{N_c} \right)  \right. \right. \cr
& \left. + N_a \left\{\left(\frac{10}{3} N_c + 2\frac{(N_c+1)(N_c-2)}{N_c} \right)\frac{N_c-2}{2} +(N_c-4) \left(\frac{10}{3}N_c + 2 \frac{N_c^2-1}{2N_c}\right)\frac{1}{2} \right. \right\}  \cr
&\left.  \left. + N_s \left\{\left(\frac{10}{3} N_c + 2\frac{(N_c-1)(N_c+2)}{N_c} \right)\frac{N_c+2}{2} +(N_c+4) \left(\frac{10}{3}N_c + 2 \frac{N_c^2-1}{2N_c}\right)\frac{1}{2} \right\} \right\} \right. \cr
&+ \left. \left(\frac{g_{i-1}^2}{(4\pi)^2} +\frac{g_{i+1}^2}{(4\pi)^2} \right) N_c N_f \frac{N_c^2-1}{2N_c}  \right] \ .
\end{align}
We do not write down the three-loop terms here, which would not fit on one page. One may derive them from the general formula in Appendix.
 As in section \ref{quiver}, there is no $K$ dependence in the beta functions at the two- (or three-) loop level. 

We look for non-trivial zeros of the beta functions by varying $N_c$, $N_f$, $N_a$ and $N_s$.
We present some examples of extremely weakly coupled fixed points together with the anomalous dimension of the permutation symmetric field strength in table \ref{table6}. 
All these examples are good candidates for dead-end CFTs. In particular, for each values of $N_c$, $N_f$, $N_a$ and $N_s$ listed there, we can choose the number of nodes $K$ in the quiver arbitrarily, so each entry on the list gives us infinitely many examples of extremely weakly coupled dead-end CFTs. It seems unlikely that the existence of these fixed points can be refuted by any other methods.

\begin{table}[htbp]
\begin{center}
\begin{tabular}{|c|c|c|c|c|c|}
\hline
 $N_c$ &$N_f$ & $N_s$ & $N_a$  &  2-loop & 3-loop  \\ \hline \hline
 29 & 1 &0  & 5  & 1.69639 $\times 10^{-5}$ &1.69638 $\times 10^{-5}$  \\ \hline
 25 & 1 & 4 & 0 & 2.26103 $\times 10^{-5}$  &2.26100  $\times 10^{-5}$  \\ \hline
17 & 1 & 1 & 4 & 4.95841 $\times 10^{-5}$   &4.95828 $\times 10^{-5}$  \\ \hline
19 & 2 & 3 & 0  & 3.80362 $\times 10^{-5}$ & 3.80356 $\times 10^{-5}$ \\ \hline 
23 & 2 & 0 & 4  & 2.61755 $\times 10^{-5}$ & 2.61752 $\times 10^{-5}$ \\ \hline 
17 & 3 & 0 & 3  & 4.66097 $\times 10^{-5}$ & 4.66088 $\times 10^{-5}$ \\ \hline 
\end{tabular}
\end{center}
\caption{ Examples of infinite series of very small anomalous dimensions of the permutation symmetric field strength based on $SU(N_c)^K$ chiral quiver gauge theories with $N_s$ generations of Bars-Yankielowicz multiplets and $N_a$ generations of Georgi-Glashow multiplets. Each entry has an additional integer label $K\ge3$.} \label{table6}
\end{table}

\section{Discussions}
In this paper, we have looked for dead-end CFTs in the perturbative regime. There are no such candidates in three dimensions, but there are infinitely many candidates in four dimensions. It would be interesting to see how far we can go beyond perturbation theory.
In two dimensions, there is an intriguing non-perturbative result reported in  \cite{Hellerman:2009bu} based on modular invariance. 
It was proved there that every unitary two-dimensional CFT with total central charge $c+\bar{c}$ contains a primary operator with dimension $\Delta_1$ that satisfies $0<\Delta_1 < \frac{c+\bar{c}}{24} + 0.473695$.
This means that as long as there is no  spin half or spin one primary operator with $\Delta<2$, the total central charge $c+\bar{c}$ must be greater than 18.270  in order to construct a dead-end CFT.\footnote{The statement as it is does not give an unconditional no-go theorem for CFTs with total central charge less than $18.270$ because the predicted operator with scaling dimensions $\Delta_1<2$ might not be a scalar.}
 A further refinement of the argument and the lower bound are found in \cite{Qualls:2013eha}\cite{Qualls:2014oea}.
In particular, \cite{Qualls:2014oea} discussed CFTs without any even spin primary operators, and within this class the total central charge must be greater than $2.227$ to construct a dead-end CFT.
In two dimensions, extremal CFTs (see e.g. \cite{Witten:2007kt} and reference therein), if they exist, give examples of dead-end CFTs. More concretely, the simplest extremal CFT is the monster CFT and its existence (therefore an example of a  non-trivial dead-end CFT) is long known \cite{Frenkel:1988xz}.

Such a bound from the central charge would be interesting in higher dimensions. We have found infinitely many candidates for dead-end CFTs in four dimensions, but the construction based on chiral gauge theories required  a large number of fields, and the infinite series we have found require  more and more matter fields.  We conjecture that there is a lower bound on the central charge (say ``a", which couples to the Euler number in the trace anomaly) that is needed to construct a dead-end CFT.\footnote{Without extra conditions, the author believes that the lowest bound for a dead-end CFT comes from the free $U(1)$ gauge theory. Unfortunately, we even do not know examples of non-free CFTs whose central charge is less than that of the free $U(1)$ gauge theory in four dimensions. To the author's knowledge, the only non-trivial candidate is the hypothetical CFT sitting at a kink of $\mathcal{N}=1$ superconformal bootstrap discussed in \cite{Poland:2011ey}. It, however, possesses a relevant deformation.}

We cannot resort to modular invariance in higher dimensions, but recent  developments in the conformal bootstrap may shed some light. In particular, the study of the energy-momentum tensor correlation functions may help. We stress again that at least in four dimensions, we do have candidates for dead-end CFTs, so a naive search without any further assumptions should be pointless. Of course, we may add a constraint on the central charges, and then the game will become non-trivial. On the other hand, it is interesting to see what the conformal bootstrap can tell us in three dimensions.

To illustrate a subtle point in the use of the conformal bootstrap here, let us take free Maxwell theory in three dimensions. By using ``electro-magnetic duality", we can show that all gauge invariant correlation functions can be mapped to  correlation functions of a free scalar field with (gauged) shift symmetry. Since the free scalar field theory with a shift symmetry forms a consistent subsector of the full CFT (given by a conformal scalar in flat space time) without any relevant scalar local operators, we may regard Maxwell theory as an example of a dead-end CFT even though the Maxwell action is not manifestly invariant under conformal symmetry (see e.g. \cite{ElShowk:2011gz}). If one only looks at the consistent subsector of the constraint from the conformal bootstrap (e.g., coming from the energy-momentum tensor four-point functions), we cannot detect any difference between the two theories although the free  conformal scalar without a shift symmetry is not an example of a dead-end CFT (because a mass term is allowed). 
In a similar manner, it is hard to exclude the possibility that global symmetry  rather than theoretical consistency forbids the appearance of relevant deformations in the conformal bootstrap approach.
In addition, it seems  more difficult to detect the possibility of adding Chern-Simons terms because the  local density is not gauge invariant.

Despite the failure of our perturbative search in three dimensions, the author believes that dead-end CFTs will exist in three dimensions, at least in the ``large central charge limit". This conviction comes from AdS/CFT. It seems that there is nothing wrong with having classical gravity in large AdS space-time without any massless or ``tachyonic" matter in $d=1+3$ dimensions. Indeed, if our universe had a tiny {\it negative} cosmological constant, the AdS/CFT dual of our universe would be a dead-end CFT because all scalar masses would be much larger than the AdS scale.
 The conformal bootstrap should be satisfied in this regime, and we would not be able to exclude it from the conformal bootstrap analysis. Recent attempts to obtain string constructions with a large gap in the spectrum may be found in \cite{Polchinski:2009ch}\cite{deAlwis:2014wia}.

We would like to end this paper with some variations of the game. Does an $\mathcal{N}=4$ supersymmetric dead-end CFT exist in four dimensions? The answer is no. The energy-momentum tensor multiplet always contains a dimension-two scalar. Does an $\mathcal{N}=2$ supersymmetric dead-end CFT in four dimensions exist? The answer is also no. The energy-momentum tensor multiplet always contains a dimension-two scalar.
How about $\mathcal{N}=1$? At this point, the energy-momentum tensor multiplet does not contain a relevant scalar operator, so there is a chance that a dead-end SCFT could exist. As a bonus, such a theory does not possess any continuous global symmetry  (except R-symmetry) because the current multiplet contains the dimension-two scalar.
However, in the Lagrangian description, one can always construct relevant deformations, such as gaugino mass terms for vector multiplets or scalar mass terms for chiral multiplets (see e.g. \cite{Kiritsis:2013gia} for a related remark), so the construction should be non-perturbative. On the other hand, pure (gauged) supergravity in $d=1+4$ dimensions can couple to only heavy matter, and it can be regarded as an AdS dual of an $\mathcal{N}=1$ dead-end SCFT.
Maybe the $\mathcal{N}=1$  variation of the game is as interesting as the one discussed in this paper.

\section*{Acknowledgements}
The author would like to thank the organizers of two wonderful workshops ``Conformal Field Theories in Higher Dimensions (Back to the Bootstrap 3)" at CERN and ``Higher Spin Symmetries and Conformal Bootstrap" at Princeton for stimulating atmospheres where he got gradually convinced to write up the material discussed here.
He would like to thank Elias Kiritsis, Luminita Mihaila and Slava Rychkov for correspondence and discussions. He also acknowledges John Schwarz  for careful reading of the manuscript. 
This work is supported by a Sherman Fairchild Senior Research Fellowship at the California Institute of Technology  and DOE grant number DE-SC0011632.

\appendix
\section{Three-loop beta functions of general multiple gauge theories}
In this appendix, we review the recent results of three-loop beta functions  of gauge coupling constants for general multiple gauge theories \cite{Mihaila:2014caa} (see also \cite{Pickering:2001aq} for a single gauge group).
We consider the direct product of simple gauge groups $G_i$ with the gauge coupling constants $g_i$ ($i= 1,\cdots, n$). For a field transforming under the representation $R$ of the gauge group $G_i$ with the generators $R^a$ in matrix notation satisfying
\begin{align}
[R^a,R^b] = i f^{abc} R^c \ ,
\end{align}
we define Casimir invariants as
\begin{align}
\mathrm{Tr}(R^a R^b) = \delta^{ab} T(R) \ , \cr
R^a R^a = \mathbf{1}_{d(R)} C(R) , \cr
f^{acd} f^{bcd} = \delta^{ab} C(G) \cr
\delta^{aa} = d(G) \ .
\end{align}
The following identity holds
\begin{align}
C(R) d(R) = T(R) d(G) \ ,
\end{align}
where $d(R)$ is the dimension of the representation $R$ and $d(G)$ is the dimension of the group. 

Explicitly, for $SU(N_c)$ we have
\begin{align}
d(G) &= N_c^2-1 \cr
C(G) &= N_c \cr
d(R) &=  \left(N_c,\frac{N_c(N_c+1)}{2},\frac{N_c(N_c-1)}{2}\right) \ \ \text{for} \ \ R= (\bar{F},S,AS) \ , \cr
T(R) &= \left(\frac{1}{2},\frac{N_c+2}{2},\frac{N_c-2}{2} \right) \cr
C(R) &= \left(\frac{N_c^2-1}{2N_c},\frac{(N_c-1)(N_c+2)}{N_c},\frac{(N_c+1)(N_c-2)}{N_c}\right) \ .
\end{align}

In general, a matter Weyl fermion $F$ in a representation $R$ is charged under multiple gauge groups. Following \cite{Mihaila:2014caa}, we use the notation $d(F_i)$ to specify the dimension of the representation $R$ with respect to the $i$th gauge group $G_i$. Furthermore, we also define the multiplicity of a representation with respect to a subset of the original direct product of simple gauge groups as
\begin{align}
D(F_i) = \prod_{ \substack{j\neq i \\ j=1}}^n d(F_j) \ , \ \ D(F_{ij}) = \prod_{\substack{k\neq i,j \\ k=1}}^n d(F_k) \ , \ \ D(F_{ijk}) = \prod_{\substack{l\neq i,j,k \\ l=1}}^n d(F_l)  \ .
\end{align}

For generic multiple gauge theories with arbitrary representations of Weyl fermions, the 
three-loop beta function of the coupling constant $g_i$  in the Modified Minimal Subtraction scheme is given by
\begin{align} 
\beta_i &= \frac{g_i^3}{(4\pi)^2} \left[-\frac{11}{3} C(G_i) + \sum_F \frac{2}{3} T(F_i)D(F_i) \right] \cr
&+\frac{g_i^3}{(4\pi)^2} \left[ \frac{g^2_i}{(4\pi)^2} \left\{-\frac{34}{3}C(G_i)^2 + \sum_F \left(\frac{10}{3} C(G_i) + 2C(F_i)\right) T(F_i)D(F_i) \right\} \right. \cr
&+ \left. \sum_{j\neq i}\frac{g_j^2}{(4\pi)^2} \sum_F 2C(F_j) d(F_j)T(F_i)D(F_{ij})  \right]  \cr
& + \frac{g_i^5}{(4\pi)^4} \left[ \frac{g_i^2}{(4\pi)^2} \left\{-\frac{2857}{54} C(G_i)^3 + \sum_F \left(\frac{1415}{54}C(G_i)^2 + \frac{205}{18}C(G_i)C(F_i)-C(F_i)^2 \right) T(F_i) D(F_i) \right. \right. \cr
&- \left. \left.  \sum_{F_m,F_n} \left(\frac{79}{54}C(G_i)+\frac{11}{9}C(F_{m,i}) \right) T(F_{m,i}) T(F_{n,i}) D(F_{m,i}) D(F_{n,i}) \right\} \right. \cr
&\left. + \sum_{j\neq i} \frac{g_j^2}{(4\pi)^2} \sum_F 2\left(2(C(G_i)-C(F_i)\right) T(F_i) C(F_j) D(F_{ij}) \right] \cr
 &+ \frac{g_i^3}{(4\pi)^2} \left[\sum_{j\neq i} \frac{g_j^4}{(4\pi)^4} \left\{ \sum_F \left(\frac{133}{18} C(G_j) -C(F_j) \right) C(F_j) T(F_i) D(F_{ij}) \right. \right. \cr
& \left. -\sum_{F_m,F_n} \frac{11}{9} C(F_{m,j}) T(F_{n,j}) T(F_{m,i}) D(F_{m,ij})D(F_{n,j}) \right\}  \cr
& + \left. \sum_{j\neq k\neq i} \frac{g_j^2}{(4\pi)^2}\frac{g_k^2}{(4\pi)^2} \left(-\sum_F C(F_j) C(F_k) T(F_i) D(F_{ijk}) \right) \right] \ . \label{betagene}
\end{align}
In our applications, there are no matter Weyl fermions that are charged under three different gauge  groups, so the last line in \eqref{betagene} will be dropped. In  \cite{Mihaila:2014caa}, one may also find the additional contributions from scalars that we do not use in this paper.

\end{document}